\documentclass[twocolumn]{aastex62}
\newcommand{\wdurl}{\href{http://www.astro.umontreal.ca/~bergeron/CoolingModels/}{\url{http://www.astro.umontreal.ca/~bergeron/CoolingModels/}}}
\usepackage{float}

\submitjournal{ApJL}

\shorttitle{The IFMR with Gaia}
\shortauthors{El-Badry et al.}

\begin{document}

\title{An Empirical Measurement of the Initial-Final Mass Relation with Gaia White Dwarfs}

\email{kelbadry@berkeley.edu}

\author{Kareem El-Badry}
\affiliation{Department of Astronomy and Theoretical Astrophysics Center, University of California Berkeley, Berkeley, CA 94720}
\affiliation{Max Planck Institute for Astronomy, D-69117 Heidelberg, Germany}

\author{Hans-Walter Rix}
\affiliation{Max Planck Institute for Astronomy, D-69117 Heidelberg, Germany}

\author{Daniel R. Weisz}
\affiliation{Department of Astronomy and Theoretical Astrophysics Center, University of California Berkeley, Berkeley, CA 94720}

\begin{abstract}
We use data from {\it Gaia} DR2 to constrain the initial-final mass relation (IFMR) for field stars with initial masses $0.9 \lesssim m_{\rm in}/M_{\odot} \lesssim 8$. Precise parallaxes have revealed unprecedented substructure in the white dwarf (WD) cooling sequence on the color-magnitude diagram (CMD). Some of this substructure stems from the diversity of WD atmospheric compositions, but the CMD remains bimodal even when only spectroscopically-confirmed DA WDs are considered. We develop a generative model to predict the CMD for DA WDs as a function of the initial mass function (IMF), stellar age distribution, and a flexibly parameterized IFMR. We then fit the CMD of 1100 bright DA WDs within 100 pc, for which atmospheric composition and completeness are well-understood. The resulting best-fit IFMR flattens at $3.5 \lesssim m_{\rm in}/M_{\odot}\lesssim 5.5$, producing a secondary peak in the WD mass distribution at $m_{\rm WD} \sim 0.8 M_{\odot}$. Our IFMR is broadly consistent with weaker constraints obtained from binaries and star clusters in previous work but represents the clearest observational evidence obtained to date of theoretically-predicted non-linearity in the IFMR. A visibly bimodal CMD is only predicted for mixed-age stellar populations: in single-age clusters, more massive WDs reach the bottom of the cooling sequence before the first lower-mass WDs appear. This may explain why bimodal cooling sequences have thus far evaded detection in cluster CMDs.

\end{abstract}
\keywords{white dwarfs
 --- stars: evolution --- Galaxy: stellar content}

\section{Introduction} 
\label{sec:intro}
The white dwarf (WD) cooling sequence in color-magnitude space encodes a wealth of information about stellar physics and Galactic evolution. WDs of different masses become dimmer and redder along parallel tracks as they age, so that if a WD's distance is known, photometry alone can strongly constrain both its mass and cooling age. The density of WDs along a cooling sequence is diagnostic of the distribution of WD ages, while the density of WDs {\it across} cooling sequences is informative about the mass distribution of WDs. For a population of a given age and initial mass function (IMF), the WD mass distribution is a strong probe of the IFMR, which relates the initial mass of star to the mass of the WD it leaves behind.

The IFMR is theoretically imperfectly understood due to difficulties in modeling the short-timescale evolution of AGB stars \citep[e.g.][]{Salaris_2009}. Most IFMRs used in the literature were measured empirically, primarily in star clusters \citep[e.g.][]{Weidemann_1983, Kalirai_2005,  Kalirai_2008, Catalan_2008, Dobbie_2012, Kalirai_2014, Cummings_2015, Raddi_2016}. Clusters offer the advantage that all stars are observed at the same distance, minimizing scatter in the CMD due to distance uncertainties. However, because all stars in a cluster have the same age and old WDs are faint, studies relying on clusters typically only probe the IFMR over the narrow range  of masses where the progenitor lifetime is of order the cluster age. The uncertainties and scatter and in observationally-inferred IFMRs remain substantial \citep{Ferrario_2005, Catalan_2008, Williams_2009}.

Field WDs can also constrain the IFMR. In the special case of WDs in wide binaries with a main sequence star, the WD's initial mass can be estimated by combining its cooling age, an age estimate of the main sequence star, and an initial mass -- pre-WD lifetime relation \citep[e.g.][]{Catalan_2008b, Zhao_2012}. More generally, the distribution of field WDs on the CMD contains information about the WD mass distribution, which yields the IFMR for a given IMF. Some studies have modeled the mass distribution of field WDs with spectroscopically estimated masses, but characterizing the completeness of observed WD samples has been a persistent challenge \citep[e.g.][]{Liebert_2005, Kepler_2007, Kleinman_2013, Limoges_2015, Tremblay_2016}. To date, most studies of the field WD population have assumed a fixed IFMR and constrained other aspects of the WD population \citep[e.g.][]{Tremblay_2014, Toonen_2017}. 

The second {\it Gaia} data release \citep{Brown_2018} provides accurate trigonometric parallaxes for a large fraction of known WDs and improves the completeness of the nearby WD population by identifying many new WD candidates. Accurate parallaxes make precise mass determination of most WDs straightforward \citep{Carrasco_2014} and reveal substructure in the CMD that has previously never been observed. In this Letter, we present constraints on the IFMR obtained from a large, homogeneous, and very nearly complete sample of field WDs on the {\it Gaia} CMD.

\section{Methods}
\label{sec:methods}

\subsection{Observational data}
\label{sec:data}

We select candidate WDs using the color and magnitudes cut described in \citet{Gaia_2018}. To minimize the effects of incompleteness, we only consider targets with \texttt{parallax\_over\_error > 20} and \texttt{parallax > 10}, corresponding to a distance limit of 100 pc. We apply the quality cuts described in Section 2.1 of \citet{Gaia_2018}, excluding the cuts on \texttt{visibility\_periods\_used} and extinction. We do not attempt to correct for extinction, which is expected to have minor effects for nearby WDs. 

Defining the absolute magnitude $\rm M_{G} = G + 5\log(\varpi) - 10$, where $\varpi$ is the parallax in mas, we only attempt to fit the CMD for objects with $\rm M_{G} < 14$, which corresponds approximately to $T_{\rm eff} \gtrsim 6000\,{\rm K}$ and a cooling age of $<2$\,Gyr. This removes about half of the {\it Gaia} WDs within 100 pc but leaves a substantially cleaner sample within which selection biases are minimized. The resulting WD sample is very nearly complete: a cut of ${\rm M_{G}} < 14$ corresponds to $\rm G<19$ at 100 pc; at this magnitude, the completeness of {\it Gaia} DR2 is $\gtrsim$\,95\% outside crowded fields \citep[see][]{Arenou_2018}. By investigating the effects of each quality cut on the candidate WD sample as a function of color and magnitude, we have verified that for WDs within 100 pc with $\rm M_{G} < 14$, none of the quality cuts we impose introduce systematic biases in color or absolute magnitude at a level greater than $\sim$2\%. 

\begin{figure*}
\includegraphics[width=\textwidth]{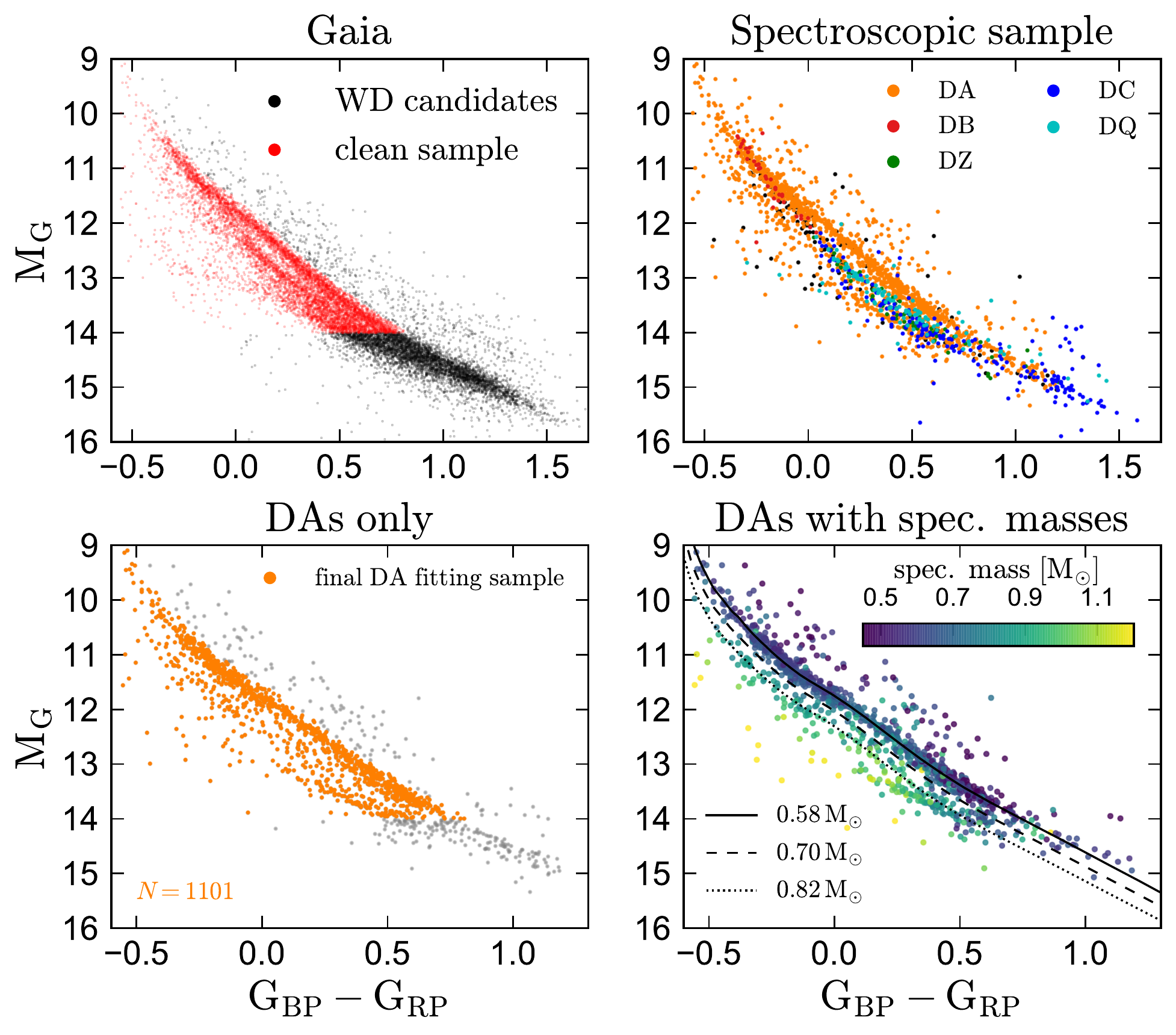}
\caption{\textbf{Top left}: {\it Gaia} WDs within 100 pc. Black points show all objects passing initial color and quality cuts. Red points show the cleaned sample, which is essentially complete at $\rm M_{G} < 14$. {\bf Top right}: Spectroscopically classified WDs within 100 pc. Some of the substructure in the CMD is due to separation of DAs and DCs/DZs/DQs. {\bf Bottom left}: Gold points show our final analysis sample. Some substructure is visible even for spectroscopically confirmed DAs. {\bf Bottom right}: Mass estimates for DAs from spectroscopic $\log g$. These are not used in our analysis but are in good agreement with the overplotted photometric tracks.}
\label{fig:wd_sample}
\end{figure*}

The upper left panel of Figure~\ref{fig:wd_sample} shows the {\it Gaia} WD sample in color-magnitude space. In addition to the cuts described above, we remove objects that fall above the primary cooling sequence. These objects are likely a combination of unresolved WD-WD binaries and undermassive WDs produced through mass transfer in binary evolution. We remove objects that fall above the cooling sequence for a DA WD with $m_{\rm WD} = 0.49 M_{\odot}$.

\subsubsection{Spectroscopic subsample}
We cross-match the 100 pc {\it Gaia} sample with the Montreal White Dwarf Database. This database is a compilation of published spectroscopic WD data, drawing in large part on SDSS spectra \citep{Dufour_2017}. The distribution of spectroscopically classified WDs in the CMD is shown in Figure~\ref{fig:wd_sample}. DC, DQ, and DZ white dwarfs are concentrated in the lower sequence of the CMD (top right panel). Due to uncertainties in the effects of convective mixing and possible additional sources of opacity in cool WDs \citep[e.g.][]{Bergeron_2001, Dufour_2005}, as well as the difficulty of obtaining independent mass estimates for WDs that are not DAs or DBs, well-calibrated cooling tracks for these objects elude us. Modeling the non-DA WDs with pure-helium atmospheres would imply that DCs, DQs, and DZs have substantially different intrinsic mass distributions from DAs and DBs \citep{Gaia_2018, Kilic_2018}, which is not expected \citep{Dufour_2005, Kleinman_2013}. 

For the remainder of this paper, we therefore only attempt to model the sample of spectroscopically confirmed DAs, subject to the cuts described above. This sample contains 1101 objects, which are shown in gold in the lower-left panel of Figure~\ref{fig:wd_sample}. Substructure in the CMD is apparent even when only DAs are considered: the lower-left panel of Figure~\ref{fig:wd_sample} shows at least two distinct sequences. Spectroscopic mass estimates for DAs (lower-right panel; not used for our analysis) are generally in good agreement with the predictions of the photometric models we use to fit the CMD.

Unlike the {\it Gaia} sample, the spectroscopic sample is not volume-complete. We estimate the spectroscopic selection function in coarse CMD bins by dividing the CMD of spectroscopically classified WDs by the {\it Gaia} CMD and then smoothing with a Gaussian filter. We note that by fitting only DAs, we are only directly measuring the IFMR for DA progenitors; or, equivalently, we are assuming that all spectral types have similar mass distributions.

\subsection{Models}
\label{sec:models}
We calculate synthetic photometry for DA WDs of a given mass and cooling age based on evolutionary model grids with carbon-oxygen cores.\footnote{Synthetic colors for these models are publicly available in some passbands are at \wdurl. Colors in the revised {\it Gaia} DR2 passbands were kindly provided by P. Bergeron.} These models are based on theoretical cooling sequences and model atmospheres calculations and are calibrated to observations of nearby WDs \citep[see][]{Fontaine_2001, Holberg_2006, Kowalski_2006, Tremblay_2011, Bergeron_2011}. ``Thick'' hydrogen atmospheres with $M_{\rm H}/M_{\star}=10^{-4}$ are assumed. We linearly interpolate between models to build a fine grid of cooling tracks for WDs of different masses. 

\subsection{Synthetic CMDs}
\label{sec:synth_cmd}

We construct synthetic CMDs and fit them to the observed data following the method outlined in \citet{Dolphin_2002}. We bin the observed CMD into a Hess diagram of 100 $\times$ 100 pixels over $9 < {\rm M_{G}}  < 14$ and $-0.6 < \rm G_{BP} - G_{RP} < 0.9$. We create synthetic Hess diagrams by binning finely spaced single-mass cooling sequences on the same grid as the data. Each sequences is weighted by the predicted WD mass distribution, which is given by
\begin{equation}
\label{eq:mass_dist}
\frac{{\rm d}N}{{\rm d}m_{{\rm WD}}}=\frac{{\rm d}N}{{\rm d}m_{{\rm in}}}\left| \frac{{\rm d}m_{{\rm WD}}}{{\rm d}m_{{\rm in}}}\right |^{-1},
\end{equation}
where ${\rm d}N/{\rm d}m_{{\rm in}}$ is the IMF and ${\rm d}m_{\rm WD}/{\rm d}m_{{\rm in}}$ is the derivative of the IFMR. We model the high-mass IMF as a power-law,  ${\rm d}N/{\rm d}m_{{\rm in}}\sim m_{\rm in}^{-\alpha}$ and calculate progenitor lifetimes from \texttt{MIST} isochrones \citep{Choi_2016}, assuming $\rm [Fe/H]=0$. Points along each cooling sequence are weighted by the age\footnote{I.e., the cooling age plus the pre-WD lifetime of a star whose initial mass corresponds to the mass of the cooling sequence.} distribution predicted in the 100 pc sample and by their time spacing. Prior to comparison with the data, the model Hess diagram is multiplied by the spectroscopic selection function in color-absolute magnitude space.

We parameterize the vertically-integrated age distribution of all stars formed as a Gaussian with free parameters $\mu_{\rm age}$ and $\sigma_{\rm age}$, truncated outside $0<{\rm age/Gyr}<12$. Limiting the observed sample to objects within 100 pc introduces a bias against old stars, which have a larger scale height than young stars. We correct for this bias following \citet[][their Equation 1]{Tremblay_2016}. We assume that the Sun is 20 pc above the Galactic midplane and that the scale height of stars of a given age is proportional to their vertical velocity dispersion, using the empiric age-velocity dispersion relation from \citet{Seabroke_2007}. This correction enhances the number density of the youngest WDs in the 100 pc sample by a factor of $\sim$2.5 relative to that of the oldest WDs, where age is measured since the formation of the progenitor star.

For WDs within 100 pc with $\rm M_{G} < 14$, typical uncertainties in color and absolute magnitude (including parallax errors) are similar, with a median $\sigma_{\rm (G_{BP}-G_{RB})} = 0.011$ and a median $\sigma_{\rm M_{G}} = 0.015$ mag. This corresponds to less than one third of the vertical pixel scale in the Hess diagram and two thirds of a the horizontal pixel scale. We only attempt to model color errors because they are larger in terms of Hess diagram pixels and because color errors broaden single-mass tracks on the CMD more than absolute magnitude errors, leading to more significant effects on the inferred IFMR.

\begin{figure*}
\includegraphics[width=\textwidth]{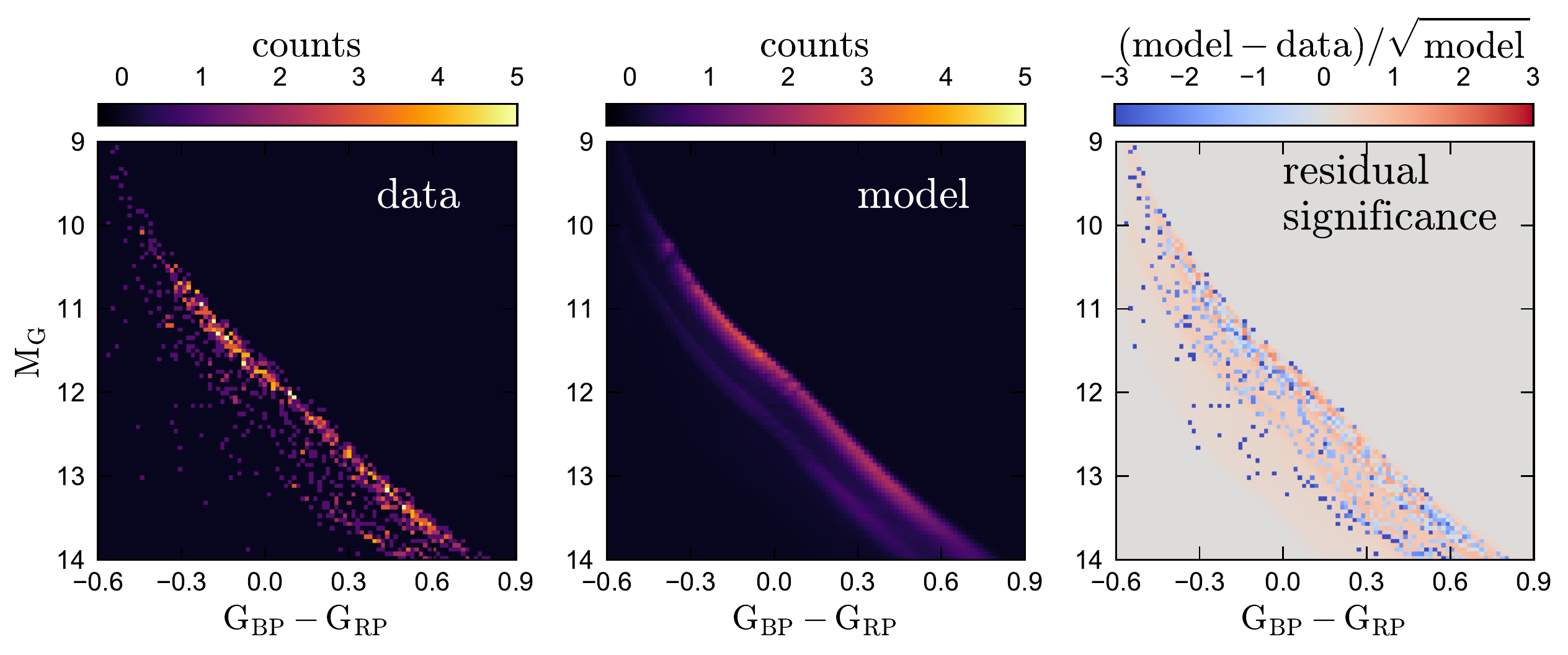}
\caption{Observed Hess diagram (left), best-fit model (middle) and residuals (right; in units of the standard deviation of the model). The model reproduces the weak bimodality of the observed population. The lower sequence in the model arises from a bimodal WD mass distribution caused by flattening in the IFMR.}
\label{fig:model_data_resid}
\end{figure*}

To account for the effects of broadening due to color uncertainties, we measure the median $\sigma_{\rm (G_{BP}-G_{RP})}$ as a function of $\rm M_{G}$ in the observed sample and convolve each horizontal row of the synthetic Hess diagram with a Gaussian kernel with dispersion equal to the median observational uncertainty at its $\rm M_{G}$. We do not account for the heteroskedasticity of the uncertainties but expect this to have little effect on our results in the systematic-limited regime relevant to bright, nearby WDs.  

Our model implicitly assumes that the IFMR is independent of metallicity. This assumption may not hold in detail \citep[e.g.][]{Meng_2008, Doherty_2015}. However, most WDs in the 100 pc sample likely formed from disk stars with a narrow range of metallicities, so we expect the effects of any systematic IFMR variations to be modest in our sample. We further assume that the IFMR has no intrinsic scatter. Previous studies place upper limits on any intrinsic scatter in the IFMR at $\sim$0.05\,M$_{\odot}$ \citep{Williams_2009, Casewell_2009}.

\subsubsection{IFMR parameterization}
\label{sec:parameterization}
We model the IFMR over $0.95 < m_{\rm in}/M_{\odot} < 8$ as a continuous piecewise-linear function. That is, we define $n_{\rm break}$ arbitrary breakpoints $(m_{{\rm in},\,i},\,m_{{\rm WD},\,i})$ at which the slope of the IFMR is allowed to change, but we leave both  $m_{{\rm in},\,i}$ and $m_{{\rm WD},\,i}$ as free parameters. This allows any deviations from linearity in the IFMR to occur where the model most strongly requires them to match the data.
We also leave the value of $m_{\rm WD}$ at $m_{\rm in}/M_{\odot}=0.95$ and 8 free, resulting in $2n_{\rm break} + 2$ free parameters. 

In the limit of large $n_{\rm break}$, this parameterization can approximate any arbitrarily complex IFMR. We adopt $n_{\rm break}=3$, resulting in an IFMR consisting of 4 connected line segments with 8 free parameters. We also tested $n_{\rm break} = 4$ and 5, finding that further increasing $n_{\rm break}$ does not substantially improve the fit. Lower value of $n_{\rm break}$ produce noticeably worse fits to the data.

\subsection{Fitting the CMD}
\label{sec:fitting}
We scale the synthetic Hess diagram such that it contains the same total counts as the observed sample. We calculate the likelihood for a particular set of model parameters by summing over all pixels in the Hess diagram, assuming that the distribution of counts in each pixel is set by a Poisson process. The log-likelihood is 
\begin{equation}
\label{eq:likelihood}
\ln L=\sum_{m_{i}\neq 0}d_{i}\ln m_{i}-m_{i}-\ln\left(d_{i}!\right),
\end{equation}
where $d_i$ and $m_i$ are the pixel values in the data and model Hess diagrams. 

We assume wide, flat priors on all model parameters except the IMF slope, $\alpha$, with the added restriction that $m_{{\rm in},\,i} < m_{{\rm in},\,i+1}$ and $m_{{\rm WD},\,i} < m_{{\rm WD},\,i+1}$. We adopt a Gaussian prior on $\alpha$ centered on 2.3 with dispersion 0.1, representative of the uncertainty in the slope of the high-mass IMF \citep[e.g.][]{Weisz_2015, Rybizki_2017}. We sample the posterior using \texttt{emcee} \citep{Foreman_Mackey_2013}, monitoring convergence using the Gelman-Rubin potential scale reduction factor \citep{Gelman_1991}. 

\section{Results}
\label{sec:results}

Figure~\ref{fig:model_data_resid} compares the observed data (left) to the synthetic Hess diagram predicted for the best-fit IFMR (middle). Residuals are shown in the right panel in units of the expected scatter for a Poisson process.
In agreement with the observed data, the model CMD contains a weak secondary cooling sequence. The bimodal tracks in the model CMD arise primarily from a bimodality in WD mass. The weight of the secondary sequence also is enhanced by the fact that the progenitors of more massive WDs formed later on average and thus are more prevalent in the 100 pc sample. 

\begin{figure}
\includegraphics[width=\columnwidth]{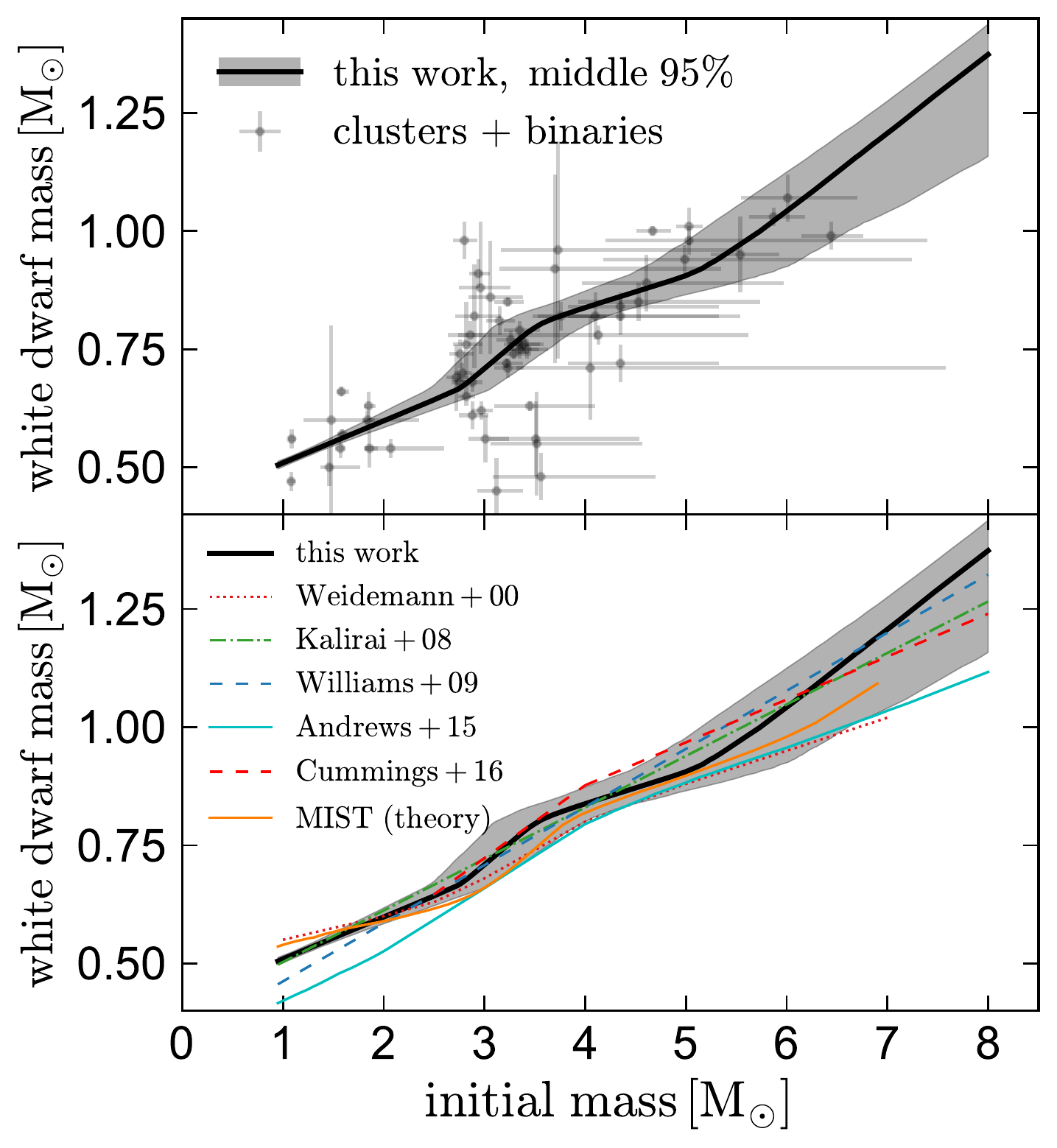}
\caption{{\bf Top}: Best-fit initial-final mass relation. Gray shaded region shows 95.4\% probability.  We parameterize the IFMR as a flexible, continuous piecewise-linear function, leaving both coordinates of the breakpoints free. Because the IFMR flattens at $3.5 \lesssim m_{\rm in}/M_{\odot} \lesssim 5.5$, stars with a wide range of initial masses accumulate at $m_{\rm WD} \sim 0.8 M_{\odot}$. Points and $1\sigma$ error bars show measurements from WDs in wide binaries and star clusters \citep{Catalan_2008}, which are shown for comparison but are not used in deriving our IFMR. {\bf Bottom}: Best-fit IFMR compared to other results from the literature \citep{Weidemann_2000, Kalirai_2008, Williams_2009,Andrews_2015, Cummings_2016, Choi_2016}.}
\label{fig:IFMR}
\end{figure}

Figure~\ref{fig:IFMR} shows the constraints on the IFMR obtained by fitting the CMD. The corresponding best-fit model parameters and their marginalized uncertainties are listed in Table~\ref{tab:ifmr}. For comparison, we show other IFMR constraints from WDs in wide binaries and clusters \citep{Catalan_2008} in the top panel. Our best-fit IFMR is generally in good agreement with these data, though the scatter between datapoints at fixed mass is substantial. {\it Gaia} constraints on the IFMR are very tight at low $m_{\rm in}$ because low-mass stars dominate the IMF, so slight changes to the IFMR at low masses cause large changes in the predicted CMD. However, the uncertainties in Figure~\ref{fig:IFMR} only represent formal fitting errors; systematics due to  uncertainties in cooling tracks and other assumptions in our model likely dominate at low masses.

\begin{table}
\centering
\caption{Best-fit IFMR parameter values and 2$\sigma$ uncertainties. The IFMR is parameterized as a piecewise linear function passing through each $(m_{\rm in}, m_{\rm WD})$ point. }
\label{tab:ifmr}
\begin{tabular}{l|l}
\hline
$m_{\rm in}/M_{\odot}$               & $m_{\rm WD}/M_{\odot}$ \\
\hline
$0.95$                     & $0.50\pm^{0.01}_{0.01}$     \\
$2.75\pm^{0.36}_{0.31}$    & $0.67\pm^{0.02}_{0.02}$  \\
$3.54\pm^{0.55}_{0.43}$    & $0.81\pm^{0.03}_{0.03}$     \\
$5.21\pm^{1.06}_{0.71}$    & $0.91\pm^{0.10}_{0.03}$     \\
$8.0$                      & $1.37\pm^{0.06}_{0.21}$ \\
\hline
\end{tabular}
\end{table}

In the lower panel of Figure~\ref{fig:IFMR}, we compare our best-fit IFMR to other parameterizations from the literature.\footnote{The plotted \citet{Cummings_2016} IFMR is their result for \texttt{PARSEC} isochrones, which are more similar to our \texttt{MIST} isochrones than their alternative \texttt{Y}$^2$ models.} The agreement with other observational studies is generally good. We note that the ``knee'' in our best-fit IFMR at $2.5 \lesssim m_{\rm in}/M_{\odot} \lesssim 4$ is predicted by stellar evolution models due to the onset of the helium flash at $m_{\rm in} \lesssim 2 M_{\odot}$ and the effects of second dredge-up at $m_{\rm in} \gtrsim 4 M_{\odot}$ \citep[e.g.][]{Dominguez_1999, Marigo_2007, Choi_2016}. The \texttt{MIST} model prediction for $\rm [Fe/H] = 0$ is shown in gold in the lower panel of Figure~\ref{fig:IFMR}. 

\section{Discussion and Conclusions}
\label{sec:conclusion}
We have shown that the distribution of nearby WDs in the CMD places strong constraints on the initial-final mass relation (IFMR), especially at initial masses $\lesssim 4M_{\odot}$. By forward-modeling the CMD for spectroscopically-confirmed DA WDs, we derive an empirical IFMR that flattens at $\sim$3.5\,M$_{\odot}$, resulting in a bimodal WD mass distribution with the usual peak at $\sim$0.58\,M$_{\odot}$ and a secondary peak at $\sim$0.8\.M$_{\odot}$ that is offset below the primary cooling sequence on the CMD. Such an IFMR is broadly consistent with previous constraints from studies of clusters and binaries. 

If the IFMR is the reason for the bimodality in the {\it Gaia} CMD of DA WDs, then the question arises why bimodal cooling sequences have not been previously identified in the WD populations of star clusters, for which uncertainties in the relative distances to individual objects are small. Aside from bimodality in some bands due to different atmospheric compositions, the CMDs of most clusters studied thus far are consistent with having tight, unimodal cooling sequences, even in studies with small photometric uncertainties \citep[e.g.][]{Hansen_2007}. 

\begin{figure}
\includegraphics[width=\columnwidth]{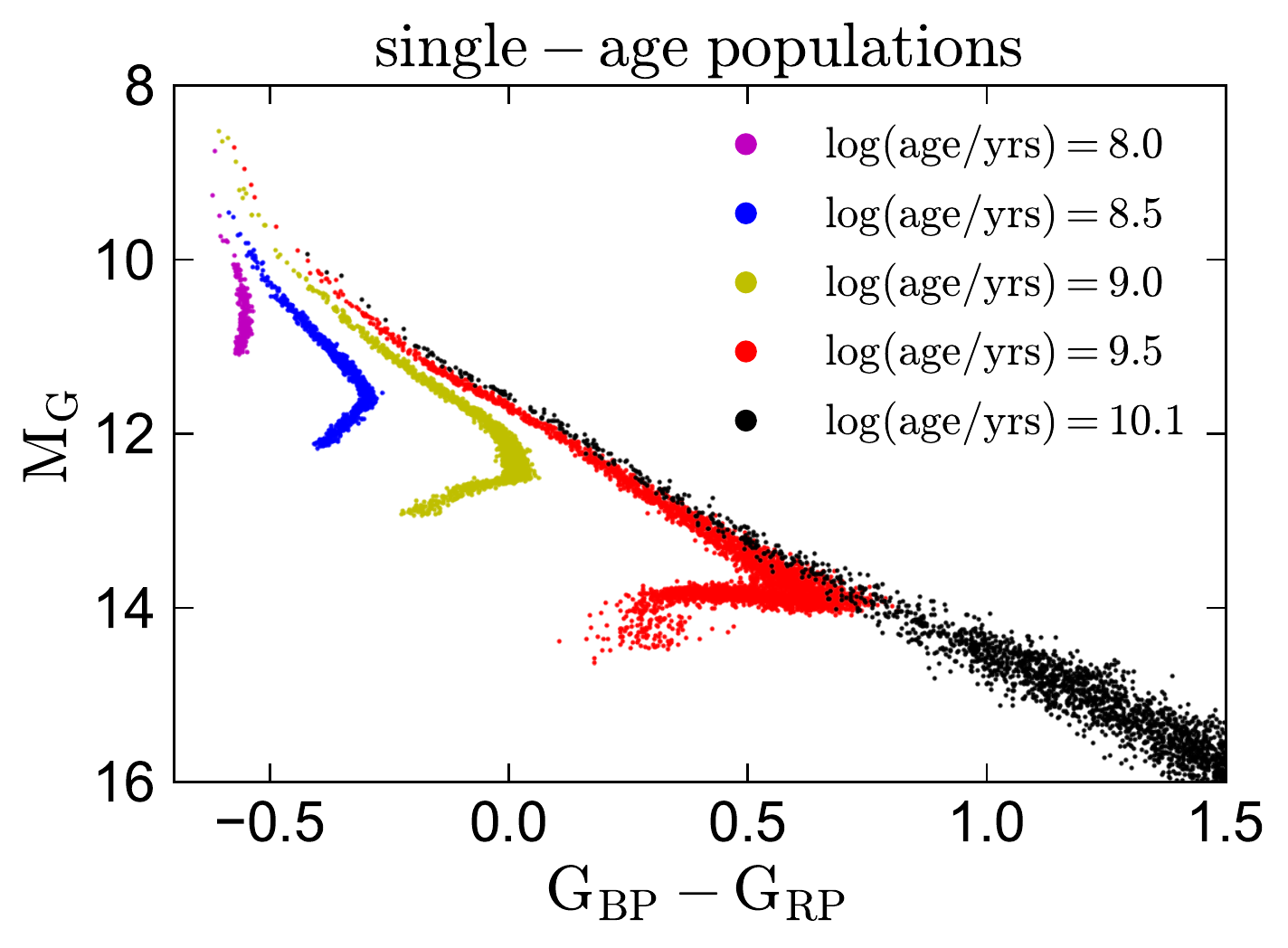}
\caption{Predicted distribution of DA white dwarfs in a single-age cluster, assuming our best-fit IFMR. Clusters are not predicted to have obviously bimodal CMDs because more massive WDs accumulate at the bottom of the cooling sequence before lower-mass WDs form. }
\label{fig:clusters}
\end{figure}

We address this question in Figure~\ref{fig:clusters}, which shows the predicted CMDs for single-age clusters, assuming the best-fit IFMR derived from the {\it Gaia} CMD. These cooling sequences show no obvious bimodality, even though the underlying total WD mass distribution is bimodal. Because the progenitors of massive WDs leave the main sequence earlier than those of lower-mass WDs, massive WDs cool and accumulate at the bottom of the cooling sequence before any lower-mass WDs appear. The WDs that most recently left the main sequence dominate the observed population in all but the nearest clusters, and these are always the lowest-mass WDs that have formed.

\subsubsection{Other possible reasons for bimodality}
\citet{Kilic_2018} recently suggested that the apparent mass bimodality visible in the {\it Gaia} CMD is due to a population of overmassive WDs formed through binary mergers. This possibility remains intriguing, though it would require the mass distribution of WD merger products to be tighter than is predicted by standard binary population synthesis models \citep[e.g.][]{Hurley_2002}. Our results do not rule out a merger-driven scenario, but they show that the observed bimodality can be naturally explained by a fairly innocuous IFMR without a significant contribution from mergers (see \citealt{Tremblay_2016} for further discussion).

Because WD mergers in single-age clusters are expected to continue to occur as a cluster ages, reheating the merger products, a merger-driven explanation for the bimodal {\it Gaia} CMD would likely also produce bimodality in the CMDs of clusters, beyond the bimodality seen in some colors due to separation of DAs and DBs. Studies of the CMDs of open clusters with sufficient photometric precision to identify two sequences, if they exist, are thus a promising route to distinguish between a merger- and IFMR-driven origin of the observed bimodality. 

\acknowledgments
We thank the referee for constructive comments and Pierre Bergeron for providing WD model photometry.
We thank Ted von Hippel, JJ Hermes, Bart Dunlap, Mukremin Kilic, David W. Hogg, Marla Geha, Eliot Quataert, Jan Rybizki, Marc Pinsonneault, and Yuan-Sen Ting for helpful discussions. 
KE was supported by the SFB 881 program (A3). 
DRW is supported by fellowships from the Alfred P. Sloan Foundation and the Alexander von Humboldt Foundation.
This research was started at the NYC Gaia DR2 Workshop at the Center for Computational Astrophysics of the Flatiron Institute in April 2018.
This work has made use of data from the European Space Agency (ESA) mission Gaia (http://www.cosmos.esa.int/gaia), processed by the Gaia Data Processing and Analysis Consortium (DPAC, http://www.cosmos.esa.int/web/gaia/dpac/consortium). Funding for the DPAC has been provided by national institutions, in particular the institutions participating in the Gaia Multilateral Agreement.

\end{document}